\documentstyle[prl,aps,twocolumn,epsfig]{revtex}
\begin{document}
\draft

\title{
Field emission from Luttinger liquids and single-wall carbon nanotubes}
\author{Alexander O. Gogolin and Andrei Komnik}
\address{Department of Mathematics, Imperial College, 180 Queen's Gate,
London SW7 2BZ, United Kingdom}
\date{Date: \today}
\maketitle
\begin{abstract}
We develop a theory for the field emission effect in Luttinger
liquids and single-wall carbon nanotubes at the level
of the energy resolved current distribution.
We generalise Fowler-Nordheim relations. Just below
the Fermi edge, we find a power-law vanishing current distribution
with the density of states exponent.
The current distribution above the Fermi edge owes its
existence to a peculiar interplay of interactions
and correlated tunnelling. It displays a non-trivial power-law
divergence just above the Fermi energy.
\end{abstract}

\narrowtext

\begin{figure}
\begin{center}
\epsfxsize=0.8\columnwidth
\hfill
\epsffile{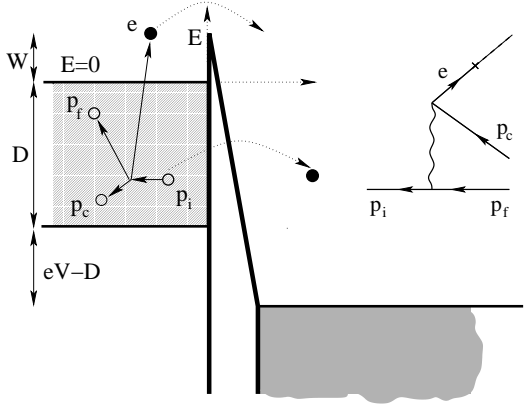}
\hfill
\vspace*{0.3cm}
\caption[]{\label{fig1}
Schematic representation of the field emission process.
Here $D$ is the conduction band-width of the emitter,
$W$ is the work function
and $V$ is the bias voltage for a related tunnelling problem (see text).
{\it In the inset:} Diagram describing the scattering of a hot hole
$p_i$ into $p_f$ and creation of an electron-hole pair $e-p_c$
with subsequent tunnelling of the electron into the right
electrode (dashed line).}
\end{center}
\end{figure}

The field emission (FE) effect has attracted considerable attention
for a long time.
Energy resolved current densities,
otherwise also termed total energy distribution functions (TEDs),
have been thoroughly studied, both experimentally
and theoretically (for a review see \cite{plummer}).
A theoretical framework for the FE was devised in the late 1920's by
Fowler and Nordheim \cite{fowler}.
They describe the emission process as electron tunnelling from a solid into
vacuum through an asymmetrical triangularly-shaped barrier.
The height of the barrier with respect to the Fermi level is given
by the work function $W$, and the slope on the vacuum side is
proportional to the electric field $F$ applied to the electrode,
as shown in Fig.~\ref{fig1}.
If the tunnelling amplitude is small,
the energy resolved current $J(\omega)$ is proportional to the
probability $n(\omega)$ for an electron to have energy $\omega$,
to the energy dependent transmission coefficient $\cal D (\omega)$,
and to a factor $\cal F$ responsible for the tip geometry:
\begin{eqnarray} \label{fnappr}
 J(\omega) = {\cal F} {\cal D}(\omega) n(\omega) \, .
\end{eqnarray}
In the noninteracting case $n(\omega)$ is given by the Fermi
distribution function.
The (quasi-classical) transmission probability
for a triangular barrier is given by \cite{landau}:
${\cal D}(\omega)\sim \exp( -4\sqrt{2m}(W-\omega)^{3/2}/3\hbar F )$,
where $m$ is the electron mass and the electron energy $\omega$ is
from now on measured relative to the Fermi energy $E_F$.
At zero temperature the emerging spectrum has a sharp threshold
at the Fermi edge (no particles with energies above the edge)
and is essentially constant in the vicinity of the edge.
Integrating over all energies one obtains the one dimensional version
of the Fowler-Nordheim formula relating the full
current to the electric field's strength \cite{fowler,plummer}.

What is even more exciting, measurements by Lea and Gomer \cite{1stexp}
revealed that in real systems there is a finite current {\it above} the Fermi
edge (`secondary current').
They found a singular TED approximately of the form
\begin{equation}
n(\omega)\sim 1/\omega
\label{fermilaw}
\end{equation}
This can only be explained by taking into account the interplay between
the electron-electron interactions and correlated tunnelling as there
is only one way for the `secondary' electron to gain additional energy.
That is by scattering on hot holes left by the `primary' electron
after tunnelling into the vacuum.
Gadzuk and Plummer applied Boltzmann-type equations for the particle
balance to explain this phenomenon \cite{gadzuk}.
Thereby they used the scattering cross-section of the process
depicted in Fig.~\ref{fig1} and obtained the TED consistent
with (\ref{fermilaw}).

In this letter we investigate the FE effect in Luttinger liquids (LLs).
The LL model is the basic model for the description of
one-dimensional (1D) interacting electrons \cite{haldane}.
It is relevant for the physics of organic quasi-1D
systems and quantum wires, for recent reviews see \cite{book,glazman}.
Our main interest is due to the fact that the LL model is
also responsible for low-energy properties of single-wall
carbon nanotubes (SWNT) as discussed in a series of papers
\cite{sammlung} over the last few years.
Especially the conductance
properties of SWNTs in various setups were exhaustively investigated both
theoretically and experimentally \cite{bockrath}.
Apart from recent widespread use of nanotubes as STM tips,
they are expected to be applied in electronics as
field emitters in high-resolution displays and cathode tubes \cite{wang}.
While there are several experimental investigations of the FE from
nanotubes \cite{french}, the theory of this process in LLs
was barely discussed in the literature so far.
In this letter we undertake to close this gap, as far as SWNTs
are concerned.
From the theoretical point of view the most interesting
part of the theory is that of the secondary current. 
To our knowledge the latter has not yet been studied experimentally
in SWNTs.
So, after a brief but comprehensive discussion of the primary current from LLs
we turn to the TED above the Fermi edge.

In order to apply Eq.~(\ref{fnappr}) to LLs one has to
know the TED $n(\omega)$ in the emitter.
In computing the primary current one neglects
higher-order tunnelling processes, so that
$n(\omega)$ is a Fourier component of the equilibrium
correlation function
$\langle \psi^\dag(t) \psi(0)\rangle$.
Here $\psi^\dag(t)$ is a local electron creation operator
at the tip of the electrode $x=0$.
The TED in this case is proportional to the local density of states
(LDOS) at the tip of the emitter.
Using the standard bosonisation scheme for open-boundary LLs \cite{book},
we write
\begin{eqnarray} \label{bosrepr}
\psi(x=0,t) = (2 \pi a_0)^{-1/2}
\exp[i \phi(x=0,t)/\sqrt{g}] \, ,
\end{eqnarray}
where $a_0$ is a lattice constant and $g$ is the LL parameter.
The gaussian chiral bosonic field $\phi(x,t)$
is governed by the LL Hamiltonian
\[
H_{LL}[\phi]=\frac{1}{4\pi}\int_{-L}^{L} dx (\partial_x\phi)^2
\]
and is periodic in $2L$, where $L$ is the system size.
(From now on we set $\hbar=v_F=e=1$.)
For a half-infinite system the LDOS's energy dependence is
known \cite{book,glazman}, so that the TED can readily be written as
\begin{eqnarray}
\label{LLraspr}
n(\omega) = \Theta(-\omega) |\omega|^{1/g-1}/a_0 D^{1/g} \Gamma(1/g) \, ,
\end{eqnarray}
where $\Gamma$ stands for the gamma function and
$D$ is the band-width of the LL.
Plugging this into Eq.~(\ref{fnappr}) one observes two important facts:
(i) at the lowest order in tunnelling the TED above the Fermi energy
is still zero, even for the interacting system,
and (ii) below the Fermi energy the TED has a power-law
singularity.
Integrating over all energies we establish the
Fowler-Nordheim-like formula for LLs:
\begin{eqnarray}\label{FN}
 J = \frac{\cal F}{a_0 D^{1/g}}
\left[ \frac{F^2}{4k_F W} \right]^{1/2g}
\exp\left( - \frac{4 k_F^{1/2} }{3F}W^{3/2} \right) \, .
\end{eqnarray}
Now we generalise these results for SWNTs.
These systems are known to be described as four-channel
LLs \cite{sammlung}.
Three channels $\phi_{c-}$ (charge-flavour), $\phi_{s+}$
(total spin), $\phi_{s-}$ (spin-flavour) are non-interacting.
The fourth channel $\phi_{c+}$ (total charge, or the plasmon
mode) possesses the LL parameter $K=(1+4U_0/\pi)^{-1/2}$, where $U_0$
is the zero Fourier component of the screened Coulomb potential.
Both the experimental \cite{bockrath} and the theoretical
\cite{sammlung} estimates place $K$ somewhere in between 0.15 and 0.3.
Note that though we now have four channels the field-operator
actually factorises as \cite{sammlung}
\begin{equation} \label{factor}
\psi\sim
\exp\{i \phi_{c+}/(4\sqrt{K})+i(\phi_{c-}+
\phi_{s+}+\phi_{s-})/4\} \, ,
\end{equation}
Therefore also the correlation functions factorise and the above results,
(\ref{LLraspr}) and (\ref{FN}), are valid for SWNTs
given the substitution
\begin{equation}
g^{-1}\rightarrow (K^{-1}+3)/4
\label{subs}
\end{equation}

In the rest of this paper we investigate the secondary current above
the Fermi edge.
The first challenge we face is to put the theory of
\cite{plummer,gadzuk} on modern footing. To do so,
let us take a broader view of the problem and
consider tunnelling between two conductors at a finite bias
voltage $V$, see Fig.~\ref{fig1}.
Clearly this becomes equivalent to the FE problem when the bias
is large, or when the energy $\omega$ of the omitted electron is
small as compared to $V$.
(From now on we concentrate on the vicinity of the Fermi
edge and neglect all trivial energy dependences of the tunnelling
probabilities.)
Allowing for finite bias is not only beneficial
technically (as it makes some integrals converge) but has important
physical consequences.
Namely, qualitative results for the secondary current will also be
valid for usual non-equilibrium tunnelling contacts between conductors
(more about it elsewhere \cite{my!}).
The model Hamiltonian for the problem can thus be taken as
\begin{equation} \label{ham0}
H = H_0[c] + H_{LL}[\psi] +
\gamma \left[ \psi^\dag(0) c(0) + c^\dag(0) \psi(0) \right] \; ,
\end{equation}
where $c^\dag(x)$ is the electron creation operator on the right of the barrier
and $\gamma$ is the effective tunnelling amplitude ($\gamma^2 \sim {\cal D}(E_F)$).
(In our approach, it is important for the electron system
on the receiving end to be non-interacting, at least asymptotically
at large distances from the barrier.)
For simplicity we use a 1D model with a single emission
channel, this will not affect the exponents.

Since we are dealing with a profoundly non-equilibrium phenomenon
we employ the Keldysh technique \cite{keldysh}.
The Keldysh Green's functions are given by
\begin{eqnarray} \label{first}
g(x,x';t-t') = -i \langle T_C \psi(x,t) \psi^\dag(x',t')
\rangle \, , \\ \nonumber
G(x,x';t-t') = -i \langle T_C c(x,t) c^\dag(x',t') \rangle \, ,
\end{eqnarray}
where
$T_C$ stands for the Keldysh contour ordering operator \cite{keldysh}.
The TED is related to the following Keldysh component
\begin{equation} \label{defEDF}
 n(0, \omega) = - i g^{-+}(0,0; \omega) \; .
\end{equation}

\begin{figure}
\begin{center}
\epsfxsize=1.0\columnwidth
\hfill
\epsffile{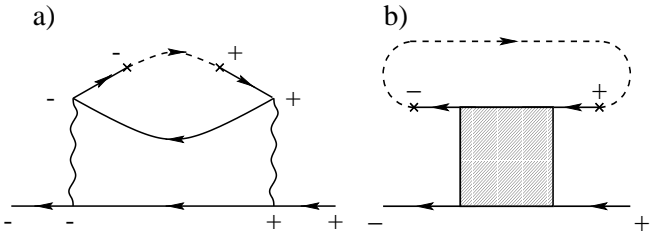}
\hfill
\vspace*{0.3cm}
\caption[]{\label{fig2}
{\it a)} The only second-order diagram contributing to the
TED above the Fermi energy.
Solid lines correspond to electrons in the emitter (LL) and the
dashed line to those on the right-hand-side of the contact.
Crosses stand for the tunnelling vertices ($\gamma$) and wiggly lines
represent the interactions ($U_0$).
{\it b)} Generic form of the diagram contributing to the TED for
$\omega>0$. The four-particle vertex ${\cal K}$ is denoted by a shaded square.}
\end{center}
\end{figure}

The zero-order perturbative expansion of the TED in the tunnelling
amplitude $\gamma$ results in already discussed Eq.~(\ref{LLraspr}).
The next non-vanishing terms are of order $\gamma^2$ and,
as pointed out in Refs.~\cite{1stexp,gadzuk},
contain the secondary emission processes.
These secondary electrons result from
inelastic scattering of hot holes, left by the primary electrons,
see Fig.~\ref{fig1}, and possess energies higher than the Fermi energy 
\cite{meden}.
There are various diagrammatic contributions to Eq.~(\ref{defEDF})
which describe such processes. However, after a closer look at the 
scattering process shown in Fig.~\ref{fig1} one realises that in the relevant
diagrams the created particles are annihilated in the same
way they were created. 
The outside electron line is then inserted into the backward
branch of the usual second order self-energy contribution.
As we do not have an {\it a priori} knowledge about how to 
decorate the vertices
with the Keldysh indices, we have checked {\it all} possibilities.
It turns out that, at this order, there is only one single
diagram that does not vanish for $\omega>0$.
It is shown in Fig.~\ref{fig2}, a) and for small positive $\omega$ equal to
$n(\omega)=CU_0^2 V/\omega$,
where $C$ is a non-universal numerical constant (related to
the suppression of the LDOS). For the FE problem we take $V\sim D$.

In real systems $\gamma$ is supposed to be small but not $U_0$.
Therefore our next objective is to develop a non-perturbative
in $U_0$ theory for the TED remaining at the $\gamma^2$ order
in the tunnelling. In the time domain the Green's function
of interest $g^{-+}(t_1,t_2)$ is given by\cite{keldysh}
\[
g^{-+}(t_1,t_2) = i \langle \psi^\dag(t_2) \psi(t_1)
\rangle = -i \langle T_C [\psi(t_1^-)\psi^\dag(t_2^+)] \rangle \; .
\]
where the superscripts indicate that
the time variables $t_1$ and $t_2$ lie on the time-ordered ($T$)
and anti-time-ordered ($\widetilde{T}$) parts of the Keldysh contour,
respectively.
A straightforward $S$-matrix expansion in powers of
$\gamma$ results in (see Fig.\ref{fig2}, b)
\[
\delta g^{-+}(t_1,t_2) = \gamma^2 \int_C \, dt_3 \, dt_4 \,
{\cal K}_C(t_1^-,t_2^+; t_3,t_4)
G(t_3,t_4) \, ,
\]
where ${\cal K}_C(t_1^-,t_2^+; t_3,t_4)=\langle T_C [
\psi^\dag(t_2^+)\psi^\dag(t_3) \psi(t_4) \psi(t_1^-)] \rangle$.
Disentangling the Keldysh indices one obtains four correlation
functions that differ in the way the field-operators are combined
under the $T$($\widetilde{T}$)-ordering operations.

The best way forward is to derive a Lehmann-type spectral representation
for the TED.
When inserting the complete set of states $|\lambda\rangle$
{\it at the break} of
the time-orderings and Fourier transforming back to the 
$\omega$-representation,
we immediately find that of the four aforementioned correlation
functions only one contributes to the TED for $\omega>0$:
\[
{\cal K}_C\to {\cal K}=  \sum_\lambda\langle 0|
\widetilde{T} [\psi^\dag(t_2)\psi^\dag(t_3)]|\lambda\rangle\langle\lambda|
T[\psi(t_1) \psi(t_4)] |0\rangle \, .
\]
Furthermore, inserting unity between the remaining field operators and
computing the time-integrals we obtain
\begin{eqnarray}
n(\omega) &=& 2 \pi \gamma^2 \sum_\mu
\Theta(V - E_\mu - \omega) |{\cal B}_\mu|^2 
\end{eqnarray}
where ${\cal B}_\mu = \sum_\nu a_{\mu \nu}a_{\nu 0}[(E_\mu-E_\nu+\omega+i0)^{-1}+(E_\nu+\omega-i0)^{-1}]$. 
Greek indices count all excited states  with
energies $E_{\nu, \lambda, \mu}$
and $a_{\mu \nu}=\langle \mu| \psi | \nu \rangle$.
Clearly all $E_\mu$'s are larger than the ground state energy $E_0$
(which we have set to zero).
Therefore we observe (though this fact is not relevant for the FE process)
that the upper threshold for the TED, at the order $\gamma^2$,
is at $\omega=V$ (physically this is clear as the hot hole
has only got so much energy to dissipate).
We can push the spectral analysis further by noticing that if the
$T$($\widetilde{T}$)-orderings in the ${\cal K}$-function were dropped
then, due to analytic properties of time integrations, the overall
result for $n(\omega>0)$ would have been zero.
Hence subtracting the un-ordered correlation function off
and assuming the LDOS to be constant
we can manipulate the time-integrals for the $\omega>0$
TED into the expression:
\begin{eqnarray} \label{com}
&~&n(\omega) = - 2i \gamma^2 \int_{-\infty}^\infty  dt
\frac{e^{i (\omega-V)t}}{t+i \alpha} \int_0^\infty d\tau_1
\int_0^\infty d\tau_2e^{i\omega(\tau_1+\tau_2)} \nonumber  \\
&~&\times
\langle
\{ \psi^\dag(\tau_1),\psi^\dag(0) \}
\{ \psi(t+\tau_1),\psi(t+\tau_1+\tau_2)\} \rangle\, .
\end{eqnarray}
where $\{.,.\}$ stands for the anti-commutator and $\alpha\sim 1/D$
is the real-time cut-off.
One advantage of this representation for the TED is that it is
explicitly vanishing in the non-interacting case
as the field operators then anti-commute at all times.
The latter does not take place in interacting systems. 
For example in LLs, representation (\ref{bosrepr}) leads to
the following {\it braiding} relationship between the field
operators, $\psi(t_1)\psi(t_2)=
e^{i\pi {\rm sign}(t_1-t_2)/g}
\psi(t_2)\psi(t_1)$.
Thus one possible interpretation of the secondary TED is that it
measures the degree of braiding induced by the interactions.

So far our treatment has been general.
Now we specialise to LLs, in which case
the four-point functions can be calculated explicitly.
Using representation (\ref{bosrepr}) we obtain:
\begin{eqnarray} \label{GF}
{\cal K}(t_1,t_2; t_3,t_4)= (2 \pi b)^{-2}
\mbox{sgn}(t_3-t_2) \mbox{sgn}(t_4-t_1) \nonumber \\
\times \Big[\frac{F(|t_3-t_2|)F(|t_4-t_1|)F^2(0)}
{F(t_2-t_1)F(t_2-t_4)F(t_3-t_1)F(t_3-t_4)}\Big]^{1/g} \, ,
\end{eqnarray}
where $F(t)=1-e^{i\epsilon_0 (t+i \alpha)}$  and $\epsilon_0=\pi/L$.
In the thermodynamic limit $L\rightarrow \infty$ we have
to expand Eq.~(\ref{GF}) in powers of $\epsilon_0$ and
retain the leading term.
The time integrals involved can be computed in terms of
generalised hypergeometric series \cite{my!} but here we only give
an integral representation convenient to work with for small energies:
\begin{eqnarray} \label{finalint}
n(\omega) = A(g,V) \int_0^{V-\omega} dE \, \frac{E^{1/g-1}}{(E+\omega)^2}
F_V(E+\omega) \, ,
\end{eqnarray}
where the spectral function is
\begin{eqnarray} \label{psiintegral}
F_V(p) = 2 \, \mbox{Im}\, e^{-i\pi/g} \int_\alpha^\infty d\xi \, \xi^{-1/g-1}
e^{-(V-p)\xi} \nonumber \\
\times \Psi^2(1/g,0,-p \xi+i\alpha) \, ,
\end{eqnarray}
and
$A(g,V) = \Gamma^2(1/g+1) \gamma^2 \alpha^{2/g} \cos^2[\pi/ 2 g]/2 \pi
a_0^2 \Gamma(1/g)$.
Here $\Psi$ stands for the Tricomi confluent hypergeometric function.
Expanding the latter on the branch cut following \cite{bateman} we obtain
the asymptotic form of the TED in vicinity of the Fermi edge
(at this final stage of the calculation we
substitute the bias voltage by $D$):
\begin{eqnarray} \label{cresult}
n(\omega) \approx C_1 (\lambda+1)^2 (\omega/D)^{\lambda} \, ,
\end{eqnarray}
where $\lambda=1/g-2$ and $C_1$ is a non-universal constant, regular at
$\lambda=-1$.
In the limit of weak interactions, $g\rightarrow 1$,
the exponent $\lambda$ approaches the Fermi liquid
result \cite{upper}.
Eq.~(\ref{cresult}) can be regarded as consisting
of two factors. The first factor is the universal $1/\omega$ divergence
inherent to all interacting systems.
The second factor reflects power-law renormalisations
normally occurring in the LLs and, in particular,
contains the LDOS of the secondary electron.
The latter object is suppressed for repulsive interactions ($g<1$).
So also the TED singularity is suppressed.
At $g_c=1/2$ the LDOS suppression effectively wins over
and the singularity disappears.

Because of the factorisation of the field operators, Eq.~(\ref{factor}),
the above results are valid for SWNTs as well. Using (\ref{subs}) we
find that for SWNT \cite{wires} $\lambda=(3+1/K)/4-2$.
Note that the critical value of the coupling now is $K_c=1/5$,
which is actually within the experimental range (see above 
Eq.~(\ref{factor})).
Therefore we do not make a specific prediction
regarding the character of the singularity.
Instead we think that both the divergent and the vanishing TEDs can be
observed depending on the experimental setup.

To conclude, in this paper we have put forward a theory
for the FE effect in LLs and SWNTs.
We have generalised Fowler-Nordheim
relations and calculated the secondary current that exhibits power-law
singularities.
We hope that this work will stimulate experiments
on the FE from SWNTs in the relevant energy range,
that is within a band $\sim 0.1 {\rm eV}$ around $E_F$.
More generally, we wish to stress
that the secondary current is a fundamental effect due to a combined
action of tunnelling and interactions.
Measurements of TED spectra for different materials could therefore reveal
important information about the nature of electron correlations
in particular media.

We are grateful to David Edwards, Reinhold Egger, Hermann Grabert and 
Georg G\"oppert
for useful discussions.
This work was partly supported by the EPSRC grant GR/N19359.


\begin{references}

\bibitem{plummer} J.~W.~Gadzuk, E.~W.~Plummer,
Rev.~Mod.~Phys. {\bf 45}, 487 (1973).

\bibitem{fowler} R.~H.~Fowler, L.~W.~Nordheim,
Proc.~Roy.~Soc.~Lond. {\bf A119}, 173 (1928).

\bibitem{landau} L.~D.~Landau, E.~M.~Lifshits,
{\it Quantum Mechanics}, Pergamon Press, Oxford (1982).

\bibitem{1stexp} C.~Lea, R.~Gomer,
Phys.~Rev.~Lett. {\bf 25}, 804 (1970).

\bibitem{gadzuk} J.~W.~Gadzuk, E.~W.~Plummer,
Phys.~Rev.~Lett. {\bf 26}, 92 (1971).

\bibitem
{haldane} F.~D.~M.~Haldane,
J.~Phys.~C: Solid State Phys. {\bf 14}, 2585 (1981).

\bibitem{book} A.~O.~Gogolin, A.~A.~Nersesyan, and A~.M.~Tsvelik,
{\it Bosonization and Strongly Correlated Systems},
Cambridge University Press (1988).

\bibitem{glazman} M.~P.~A.~Fisher, L.~I.~Glazman: in \emph{Mesoscopic Electron Transport}, edited by L.~Sohn \emph{et al.}, Kluwer Academic Publishers (1997).

\bibitem{sammlung}
R.~Egger, A.~O.~Gogolin, Phys.~Rev.~Lett. {\bf 79},  5082  (1997);
Eur.~Phys.~J.~B {\bf 3}, 281 (1998);
C.~L.~Kane, L.~Balents, M.~P.~A.~Fisher, Phys.~Rev.~Lett. {\bf 79}, 5086
(1997).


\bibitem{bockrath} S.~J.~Tans, M.~H.~Devoret \emph{et al.}, Nature {\bf 386},
474  (1997); 
M.~Bockrath, D.~H.~Cobden, J.~Lu, A.~G.~Rinzler,
R.~E.~Smalley, L.~Balents, P.~L.~McEuen, Nature {\bf 397},  598  (1999).

\bibitem{wang} Q.~H.~Wang, A.~A.~Setlur \emph{et al.}, Appl.~Phys.~Lett. {\bf 72}, 2912 (1998); Y.~Saito, K.~Hamaguchi \emph{ et al.}, Appl.~Phys. A {\emph 67}, 95 (1998).


\bibitem{french} J.-M.~Bonard, J.-P.~Salvetat \emph{et al.},
Phys.~Rev.~Lett. {\bf 81}, 1441 (1998);
J.-M.~Bonard, J.-P.~Salvetat \emph{et al.},
Appl.~Phys. A 69, 245 (1999), and references therein.

\bibitem{my!} A.~Komnik and A.~O.~Gogolin, preprint in preparation.

\bibitem{keldysh} L.~V.~Keldysh,
Zh.~Eksp.~Teor.~Fiz., {\bf 47}, 1515 (1964)
[Sov. Phys. JETP {\bf 20}, 1018 (1965)],
E.~M.~Lifshitz, L.~P.~Pitaevskii,
{\it Physical Kinetics},
Pergamon Press, Oxford (1981).
We follow the notation of the latter.

\bibitem{meden} This problem is akin to but different from that of 
hot electron relaxation, see V.~Meden, C.~Wohler, J.~Fricke \emph{et al.}, Phys.~Rev.~B 
{\bf 52}, 5624 (1995).  


\bibitem{bateman} A.~Erd\'elyi (ed.),
{\it Higher transcendental functions},
McGraw-Hill, New York (1953).

\bibitem{upper} The behavior of the TED next to the upper
threshold $\omega=V$ can also be extracted from
\protect Eqs.~(\ref{finalint}) \protect  and
\protect (\ref{psiintegral})\protect and is given by another power law,
$n(\omega) \sim (V-\omega)^{\nu}$,
where $\nu=1/g$.

\bibitem{wires} For systems like quatum wires we are dealing
with a spinful LL model.
Since the tunneling process conserves spin, the exponent changes
to $\lambda = (1/g_c + 1/g_s)/2 - 2$, where $g_{c,s}$ denote
interaction constants in charge and spin channel, respectively.
The relation between the lower- and upper threshold
exponents $\nu = \lambda + 2$ holds generally.

\end{references}
\end{document}